\documentstyle[amssymb,preprint,aps]{revtex}

\begin{document}
\title{Temperature effects on edge-state properties in the integer quantum
Hall regime}
\author{I. O. Baleva}
\address{Institute of Semiconductor Physics, National Academy of Sciences,\\
45 Pr. Nauky, Kiev 252650, Ukraine }
\author{Nelson Studart}
\address{Departmento de Fisica, Universidade Federal\\
de S\~{a}o Carlos, 13565-905, S\~{a}o Carlos, S\~{a}o Paulo, Brazil }
\author{O. G. Balev}
\address{Departmento de Fisica, Universidade Federal\\
de S\~{a}o Carlos, 13565-905, S\~{a}o Carlos, S\~{a}o Paulo, Brazil \\
and Institute of Semiconductor Physics, National Academy of\\
Sciences, 45 Pr. Nauky, Kiev 252650, Ukraine}
\date{\today}
\maketitle

\begin{abstract}
The edge and bulk structure of Landau levels (LLs) in a wide channel at the $%
\nu =1$ quantum Hall regime is calculated for not-too-low temperatures, $%
\hbar \omega _{c} \gg k_{B}T\gg \hbar v_{g}/2\ell _{0}$, where $v_{g}$ is
the group velocity of the edge states and $\ell _{0}=\sqrt{\hbar c/|e|B}$ is
the magnetic length. Edge-states correlations essentially modify the spatial
behavior of the lowest spin-up LL, which is occupied, compared to the lowest
spin-down LL, which is empty. The influence of many-body interactions on the
spatially inhomogeneous spin-splitting between the two lowest LLs is studied
within the generalized local density approximation. Temperature effects on
the enhanced spin-splitting, the position of the Fermi level within the
exchange enhanced gap and the renormalization of edge-states group velocity
by edge states screening are considered. It is shown that the maximum
activation energy $G$ in the bulk of the channel is determined by the gap
between the Fermi level and the bottom of the spin-down LL, because the gap
between the Fermi level and the spin-up LL is much larger. For the maximum
value of $G$, it is shown that the renormalized group velocity $v_{g}\propto
T$ for $T\rightarrow 0$ and, in particular, the condition of not-too-low $T$
can be satisfied for $4.2\agt T\agt0.3$ K. In other words, the regime of
not-too-low temperatures regime can be achieved even for rather low $T$. \ \ 
\end{abstract}

\pacs{73.43.-f; 73.63.Nm}

A number of works have been devoted to study the influence of Coulomb
interactions on the edge states of a two-dimensional electron system (2DES)
in the presence of a strong magnetic field.\cite%
{chklovskii92,dempsey93,brey93,suzuki93,gelfand94,balev97} Quite recently,%
\cite{balev01} the spatial behavior of the Landau levels (LL's) for a wide
channel at the filling factor $\nu =1$ was studied in the limit of very
low-temperatures, $T\ll \hbar v_{g}^{(0)}/k_{B}\ell _{0}$, using the
generalized local density approximation (GLDA) in which exchange and
correlation effects, especially due to the edge states, are included. $%
v_{g}^{(0)}$ is the group velocity of the edge states and the magnetic
length $\ell _{0}=\sqrt{\hbar /m^{\ast }\omega _{c}}$ with $\omega
_{c}=|e|B/m^{\ast }c$. Many-body effects were calculated in the screened
Hartree-Fock approximation. It was shown that the maximum activation energy
in the bulk of the channel corresponds to a highly asymmetric position of
the Fermi level within the gap between spin-down and spin-up LL's. The
renormalized edge-state group velocity $v_{g}^{(0)}$ independent on $T$
exhibits a strong decrease due to correlation effects.

In this work, we extend the previous approach of Ref. \cite{balev01} for
not-too-low temperatures, where $\hbar \omega _{c}\gg k_{B}T\gg \hbar
v_{g}/2\ell _{0}$ and the characteristic length $\ell _{T}=\ell
_{0}^{2}k_{B}T/\hbar v_{g}\gg \ell _{0}/2$ becomes relevant.\cite{balev00}
In particular, we study here the temperature effects on the enhanced
spin-splitting for the 2DES in a wide channel within the $\nu =1$ quantum
Hall regime, the position of the Fermi level within the pertinent {\it %
exchange enhanced} gap and the renormalized edge-states group velocity $%
v_{g} $, which can be strongly dependent on $T$.

We consider a wide symmetric channel along the $x$ axis in strict two
dimensions with a strong uniform magnetic field $B$ in the $z$ direction. In
the absence of exchange and correlation effects, we take the confining
potential $V_{w}(y)=0$ in the inner part of the channel and $%
V_{w}(y)=m^{\ast }\Omega ^{2}(y-y_{r})^{2}/2$ at the right edge, $y\geq
y_{r} $, and $\Omega $ is the confining frequency. We assume that $V_{w}(y)$
is smooth on the scale of the magnetic length $\ell _{0}$; i.e., $\Omega \ll
\omega _{c}$. Extending the GLDA for $T=0$ of Ref. \cite{balev01} on
not-too-low $T$, in particular, by using some results of Ref. \cite{balev00}%
, it follows that the energy spectrum $E_{0,k_{x},1}$ of lowest spin-up LL ($%
n=0,\sigma =1$) of the interacting 2DES for $r_{0}=e^{2}/(\varepsilon \ell
_{0}\hbar \omega _{c})\lesssim 1$ can be obtained from the solution of the
single-particle Schr\"{o}dinger equation with the Hamiltonian ${\cal H}=%
{\cal H}^{0}+V_{xc}(y)$, where a self-consistent exchange-correlation
potential, smooth on $\ell _{0}$ scale, is given by%
\begin{equation}
V_{xc}(y)=E_{0,y/\ell _{0}^{2},1}-\left( \frac{\hbar \omega _{c}}{2}-\frac{%
\left| g_{0}\right| \mu _{B}B}{2}+V_{w}(y)\right) .  \label{2}
\end{equation}%
Here $E_{0,k_{x},1}=\varepsilon _{0,k_{x},1}+\varepsilon _{0,k_{x},1}^{xc}$,
where $\varepsilon _{n,k_{x},\sigma }$ are the eigenvalues of the
one-electron Hamiltonian ${\cal H}^{0}=[(\hat{p}_{x}+eBy/c)^{2}+\hat{p}%
_{y}^{2}]/2m^{\ast }+V_{w}(y)+g_{0}\mu _{B}\hat{S}_{z}B/2$ and $\varepsilon
_{0,k_{x},1}^{xc}$ is the exchange and correlation contributions in the
screened Hartree-Fock approximation; $g_{0}$ the bare g-factor, $\mu _{B}$
the Bohr magneton and $\hat{S}_{z}$ is the $z$ component of the spin
operator with eigenvalues $\sigma =1$ and $\sigma =-1$ for spin up and down; 
$\varepsilon $ is the background dielectric constant. Further, Eq. (\ref{2})
is essentially different from Eq. (17) of Ref. \cite{balev01} as for $%
E_{0,k_{x},1}$ now we have a very different expression, $k_{x}\geq 0$ and $%
\tilde{v}_{g}=(\pi \hbar \varepsilon /e^{2})v_{g}$, as%
\begin{eqnarray}
E_{0,k_{x},1} &=&\frac{\hbar \omega _{c}}{2}-\frac{|g_{0}|\mu _{B}B}{2}+%
\frac{m^{\ast }\Omega ^{2}\ell _{0}^{4}}{2}(k_{x}-k_{r})^{2}\Theta
(k_{x}-k_{r})-\sqrt{\frac{\pi }{2}}\frac{e^{2}}{\varepsilon \ell _{0}}k_{%
\text{bulk}}(r_{0})f_{0,k_{x},1}+\frac{2e^{2}}{\pi \varepsilon \tilde{v}_{g}}%
f_{0,k_{x},1}  \nonumber \\
&&\times \int_{0}^{\infty }dq_{x}\exp (-q_{x}^{2}\ell
_{0}^{2})F_{1}(q_{x},k_{x}-k_{r0}^{(1)})F_{2}(q_{x},k_{x}-k_{r0}^{(1)})\left[
1+\frac{1}{\tilde{v}_{g}}\exp (-q_{x}^{2}\ell _{0}^{2}/2)r_{00}^{s}(q_{x})%
\right] ^{-1},  \label{3a}
\end{eqnarray}%
where $f_{0,k_{x},1}=1/\left[ 1+\exp (E_{0,k_{x},1}-E_{F})/k_{B}T\right]
\approx \{1-\tanh \left[ \ell _{0}^{2}(k_{x}-k_{r0}^{(1)})/2\ell _{T}\right]
\}/2,$ with $E_{F}=E_{0,k_{r0}^{(1)},1}$ being the Fermi energy renormalized
both by exchange and correlations and $\Theta (k_{x}-k_{r})$ is the step
function. The edge of the ($n=0,\sigma =1$) LL is denoted by $%
y_{r0}^{(1)}=\ell _{0}^{2}k_{r0}^{(1)}$, with $%
k_{r0}^{(1)}=k_{r}+k_{e}^{0,1} $ and $k_{r}=y_{r}/\ell _{0}^{2}$ and it is
assumed that the Fermi wave vectors $k_{r0}^{(1)}$ and $k_{e}^{0,1}$ are not
changed by exchange correlation effects, even though the Fermi energy in the
Hartree approximation $E_{F}^{H}$ is essentially different from $E_{F}$. The
fourth term of Eq. (\ref{3a}) comes from the exchange interaction, where the
additional factor $k_{\text{bulk}}(r_{0})<1$ takes into account the
decreasing of the many-body contribution due to the weak ``bulk'' screening
of the fully occupied LL caused by inter-LL virtual transitions. For assumed
conditions $k_{\text{bulk}}(r_{0})$ is well approximated by a factor
calculated exactly in Ref. \cite{balev01}, for the bulk of the channel, For
instance, we find $k_{\text{bulk}}(r_{0})$ are $0.79,0.74,0.70,0.66$ and $%
0.63$ for $r_{0}=0.6,0.8,1.0,1.2,$ and $1.4$, respectively. For $r_{0}<<1,$ $%
k_{\text{bulk}}(r_{0})$ tends to 1 and the well-known exchange contribution
is recovered. In the last term of Eq. (\ref{3a}), the function $%
r_{00}^{s}(q_{x})$, dependent on $\ell _{T}$, is presented in Ref. \cite%
{balev00} and the functions $F_{1}$ and $F_{2}$ are given by

\begin{equation}
F_{1}(q_{x},k_{x}-k_{r0}^{(1)})=\int_{0}^{\infty }dq_{y}\frac{\exp
(-q_{y}^{2}\ell _{0}^{2}/4)}{\sqrt{q_{x}^{2}+q_{y}^{2}}}\frac{\cos
\{q_{y}(k_{x}-k_{r0}^{(1)})\ell _{0}^{2}\}}{1+q_{y}^{2}\ell _{T}^{2}},
\label{4}
\end{equation}%
and 
\begin{equation}
F_{2}(q_{x},k_{x}-k_{r0}^{(1)})=\int_{0}^{\infty }dq_{y}\frac{\exp
(-q_{y}^{2}\ell _{0}^{2}/4)}{\sqrt{q_{x}^{2}+q_{y}^{2}}}\cos
\{q_{y}(k_{x}-k_{r0}^{(1)})\ell _{0}^{2}\}F(q_{y}),  \label{4a}
\end{equation}%
where $F(q_{y})=\pi \left| q_{y}\right| \ell _{T}/\sinh (\pi \left|
q_{y}\right| \ell _{T})$. The group velocity of the edge states $%
v_{g}=(\partial E_{0,k_{x},1}/\hbar \partial k_{x})_{k_{x}=k_{r0}^{(1)}},$
renormalized by exchange and correlations, is determined by a positive
solution of the equation

\begin{equation}
\tilde{v}_{g}=\tilde{v}_{g}^{H}+\frac{\pi }{4}\frac{\ell _{0}}{\ell _{T}}%
\left\{ \sqrt{\frac{\pi }{2}}k_{\text{bulk}}(r_{0})-\frac{2\ell _{0}}{\pi }%
\int_{0}^{\infty }dq_{x}\frac{\exp (-q_{x}^{2}\ell
_{0}^{2})F_{1}(q_{x},0)F_{2}(q_{x},0)}{\tilde{v}_{g}+\exp (-q_{x}^{2}\ell
_{0}^{2}/2)r_{00}^{s}(q_{x})}\right\} ,  \label{5a}
\end{equation}%
where $\tilde{v}_{g}^{H}=(\pi \hbar \varepsilon /e^{2})v_{g}^{H}$ and $%
v_{g}^{H}=v_{g0}^{1,H}$, with $v_{g0}^{1,H}=c{\cal E}_{e0}^{(1)}/B$ being
the group velocity in the Hartree approximation. Here ${\cal E}%
_{e0}^{(1)}=\Omega \sqrt{2m^{\ast }\Delta _{F0}^{(1)}}/|e|$ is the electric
field associated with the confining potential $V_{w}(y)$ at $y=y_{r0}^{(1)}$
and $\Delta _{F0}^{(1)}=E_{F}^{H}-\hbar \omega _{c}/2-g_{0}\mu _{B}B/2$.
Equation (\ref{5a}) was evaluated using $[\partial
F_{1,2}(q_{x},k_{x}-k_{r0}^{(1)})/\partial k_{x}]_{k_{x}=k_{r0}^{(1)}}=0$
and the relation $[-\partial f_{0,k_{x},1}/\partial
k_{x}]_{k_{x}=k_{r0}^{(1)}}=\ell _{0}^{2}/4\ell _{T}$. Note that since $\ell
_{T}=\ell _{0}^{2}k_{B}T/\hbar v_{g}$, we have that $\ell _{T}\propto \tilde{%
v}_{g}^{-1}$ in Eqs. (\ref{3a})-(\ref{5a}). Equations (\ref{3a}) and (\ref%
{5a}) provide the self-consistent scheme to calculate the LL spectrum in the
GLDA for not too-low temperatures. We observe that for $\tilde{v}%
_{g}^{H}\rightarrow 0$, the proper solution $\tilde{v}_{g}$ tends to a
finite value. This result is quite different from the case of very low
temperatures in which $v_{g}\propto \sqrt{v_{g}^{H}}$, for $%
v_{g}^{H}\rightarrow 0.$\cite{balev01}. If correlations due to edge-states
screening are neglected (excluding the term with integral) in Eq. (\ref{5a}%
), then it follows that for many experimentally realistic conditions Eq. (%
\ref{5a}) does not have any positive solution, i.e., the condition $\sqrt{%
\pi /32}\,r_{0}\,k_{\text{bulk}}(r_{0})[\hbar \omega _{c}/k_{B}T]<1$ is not
satisfied.

The positive gap between the bottom of the upper spin-split LL and the Fermi
level of the interacting 2DES, $G(v_{g}^{H})=E_{0,0,-1}-E_{0,k_{r0}^{(1)},1}$%
, is then written as

\begin{eqnarray}
G &=&|g_{0}|\mu _{B}B-\frac{m^{\ast }\omega _{c}^{2}}{2\Omega ^{2}}%
(v_{g}^{H})^{2}+\frac{1}{2}\sqrt{\frac{\pi }{2}}\frac{e^{2}}{\varepsilon
\ell _{0}}k_{\text{bulk}}(r_{0})-\frac{e^{2}}{\pi \varepsilon \tilde{v}_{g}}%
\int_{0}^{\infty }dq_{x}\exp (-q_{x}^{2}\ell _{0}^{2})  \nonumber \\
&&\times F_{1}(q_{x},0)F_{2}(q_{x},0)\left[ 1+\frac{1}{\tilde{v}_{g}}\exp
(-q_{x}^{2}\ell _{0}^{2}/2)r_{00}^{s}(q_{x})\right] ^{-1},  \label{6a}
\end{eqnarray}%
where $\tilde{v}_{g}\equiv \tilde{v}_{g}(v_{g}^{H})\geq 0$ is the function
of $v_{g}^{H}$ obtained from Eq. (\ref{5a}). In addition, in Eq. (\ref{6a})
it was used for the term $\propto (k_{r0}^{(1)}-k_{r})^{2}$ that $%
k_{e}^{0,1}=(m^{\ast }\omega _{c}^{2}/\hbar \Omega ^{2})v_{g}^{H}$. In the
bulk of the channel, the total gap between the lowest spin-split LLs is $%
G_{-1,1}=E_{0,0,1}-E_{0,0,-1}\approx |g_{0}|\mu _{B}B+\sqrt{\pi /2}%
(e^{2}/\varepsilon \ell _{0})k_{\text{bulk}}(r_{0})$.

In the absence of many-body interactions, the maximum value of the
dimensionless activation gap $G_{a}(v_{g}^{H})=G/(\hbar \omega
_{c}/2)\approx 0.015$. We use the parameters of GaAs based samples, in
particular $\varepsilon \approx 12.5$, $g_{0}=-0.44$ and $m^{\ast
}=0.067m_{0}$. Then the activation gap is enhanced when $G_{a}>0.015$. The
asymmetry of the Fermi level position within the Fermi gap in the bulk of
the channel can be characterized by another dimensionless function $\delta
G(v_{g}^{H})=(\bar{G}_{-1,1}-G_{a})/G_{a}$, where $\bar{G}%
_{-1,1}=G_{-1,1}/(\hbar \omega _{c}/2)$. Notice that, when $%
v_{g}^{H}\rightarrow 0$, $E_{F}^{H}$ tends, from the upper side, to the
bottom of the ($n=0,\sigma =1$) LL, in the absence of many-body interactions.

In Fig. 1 $\tilde{v}_{g}$ is depicted as a function of $\tilde{v}_{g}^{H}$
using Eq. (\ref{5a}). The solid, dashed and dotted curves correspond to $%
\hbar \omega _{c}/k_{B}T=10$, $15$ and $74.5$ or $T\approx 31.3$, $20.8$ and 
$4.2$ K respectively for $B=15.7$ T and electron density $n_{s}=3.8\times
10^{11}\;$cm$^{-2}$, $r_{0}=0.66$, $k_{\text{bulk}}(r_{0})=0.77$. From Fig.
1, it is seen that for $\tilde{v}_{g}^{H}\rightarrow 0$ the renormalized
group velocity $\tilde{v}_{g}$ tend to a finite value in contrast with the
regime of very low temperatures. We point out that for these parameters
there is no positive solution $v_{g}$ if edge-state correlations are
neglected. Notice that for $\tilde{v}_{g}^{H}=0$, depending on $v_{g}$, the
ratio $\ell _{T}/\ell _{0}=7.3$, $3.6$ and $2.0$ for the solid, dashed and
dotted curves in Fig. 1. So the assumed condition $2 \ell_{T}/\ell _{0}\gg 1$
is well satisfied in Fig. 1.

In Fig. 2, using Eqs. (\ref{5a}) and Eq. (\ref{6a}), we plot $G_{a}$ and ($%
\delta G/10)$ as a function of $\tilde{v}_{g}^{H}$. In the solid and dashed
curves we used the same parameters as in Fig. 1. $G_{a}$ is represented by
the curves with a negative slope while the curves with positive slope
represent $(\delta G/10)$. Here we use the value of the confining frequency $%
\Omega =8.2\times 10^{11}\;$s$^{-1}$, which is slight larger than $\Omega
=7.8\times 10^{11}\;$s$^{-1}$ obtained in Ref. \cite{muller92}. In this case 
$\omega _{c}/\Omega =50$. We observe that the maximum of $G_{a}$ and the
corresponding minimum of $\delta G$ are kept for $\tilde{v}%
_{g}^{H}\rightarrow 0$. In particular, in the solid curves, these values are 
$G_{a}\approx 0.43$ and $\delta G\approx 2.0$. In the dashed curves, the
values are $G_{a}\approx 0.30$ and $\delta G\approx 3.4$ as $\tilde{v}%
_{g}^{H}\rightarrow 0$. Notice that for $\tilde{v}_{g}^{H}\rightarrow 0$,
and $\tilde{v}_{g}$ given by the dotted curve in Fig. 1, we have $%
G_{a}\approx 8.7\times 10^{-2}$ and $\delta G\approx 14.1$, which gives $%
G\approx 13.5$ K, i.e. about three times greater than, the liquid helium
temperature, $T=4.2$ K. So the Fermi level is substantially closer to the
bottom of the upper spin-split LL then to the bottom of the lowest
spin-split LL. Such an asymmetry increases as $T$ decreases. Furthermore,
one can see that $G_{a}$ drops fast by increasing $\tilde{v}_{g}^{H}$. Note,
however, that for sufficiently small $G_{a}$, such that $G=(\hbar
\omega_{c}/2)G_{a}\leq k_{B}T$, our formulas, of course, are not valid.

In Fig. 3 $\tilde{v}_{g}$ is shown as a function of $T$ by solid curves for $%
\tilde{v}_{g}^{H}=0$, calculated from Eq. (\ref{5a}). The dashed curves
represent $\ell _{T}/\ell _{0}$ pertinent to the solid curves. The long
curves correspond to the parameters $n_{s}=3.8\times 10^{11}\;$cm$^{-2}$, $%
B=15.7$ T, $r_{0}=0.66$, $k_{\text{bulk}}(r_{0})=0.77$. The short curves
correspond to $n_{s}=1.26\times 10^{11}\;$cm$^{-2}$, $B=5.2$ T, $r_{0}=1.146$%
, $k_{\text{bulk}}(r_{0})=0.67$. In Fig. 3 the curves are plotted for the
temperature range satisfying $\hbar \omega _{c}/k_{B}T\geq 10$. As $T$ drops
to below $4.2$ K (as well for $T\ll 1$ K), the upper and lower dashed curves
tend to $2\ell _{T}/\ell _{0}\approx 5.2$ and $3.8$, respectively. From this
result it follows that the regime of not-too-low temperatures, $k_{B}T\gg
\hbar v_{g}/\ell _{0}$, can hold even for $T\alt1$K.

In Fig. 4 we show the energy spectra as a function of $Y=\ell
_{0}^{2}(k_{x}-k_{r0}^{(1)})/\ell _{T}$ for $\tilde{v}_{g}^{H}=0$ and the
parameters of solid curves in Figs. 1 and 2. In particular, we use $\hbar
\omega _{c}/k_{B}T=10$, $B=15.7$ T, $\omega _{c}/\Omega =50$ and $\tilde{v}%
_{g}=0.065$; here $k_{r0}^{(1)}=k_{r}$ as $\tilde{v}_{g}^{H}=0$. Using Eq. (%
\ref{3a}), $E_{0,k_{x},1}$ is indicated by the lower solid curve. The upper
solid curve represents $E_{0,k_{x},-1}=\varepsilon_{0,k_{x},-1}$ and the
dashed horizontal line is the Fermi level of the interacting 2DES.

In conclusion, we have shown that edge-state correlations modify drastically
the ($n=0,\sigma =1)$ LL spectrum of the 2DES in the $\nu =1$ quantum Hall
regime at not-too-low temperatures, $\hbar \omega _{c}\gg k_{B}T\gg \hbar
v_{g}/2\ell _{0}$, in a wide region ($\gg \ell _{T}$) nearby the channel
edge. The spectrum is obtained in a well justified self-consistent GLDA. We
have shown that the position of the Fermi level $E_{F}$ is highly asymmetric
within the gap defined by the ($n=0,\sigma =1$) and ($n=0,\sigma =-1$) LLs
in the bulk of the channel due to such correlations. Hence, the activation
gap $\Delta _{ac}=2G$ (obtained from the resistivity as $\rho _{xx}\propto
\exp (-\Delta _{ac}/2k_{B}T)$) is much smaller than the Fermi gap. In
particular, we have obtained the edge-state group velocity $v_{g}$,
renormalized by exchange and correlations effects, and an analytical
expression for $G$, the gap between the bottom of the upper spin-split LL
and $E_{F}$. We have obtained that there is a maximum of $G$ for $\tilde{v}%
_{g}^{H}=0$, which essentially depends on $T$. Moreover, our self-consistent
study shows that for $\tilde{v}_{g}^{H}=0$ the regime of not-too-low
temperatures can be achieved for very small temperatures, e.g., $4.2\agt T%
\agt0.3$ K, as for $T\alt1$ it follows that $v_{g}\propto T$ and $2\ell
_{T}/\ell _{0}$ becomes temperature independent while it is still large, $%
2\ell _{T}/\ell _{0}\gg 1$.

\acknowledgements

This work was supported by the Funda\c{c}\~{a}o de Amparo \`{a} Pesquisa de
os Estado de S\~{a}o Paulo (FAPESP)\ and the Conselho Nacional de
Desenvolvimento Cient\'{\i}fico e Tecnol\'{o}gico (CNPq).

\bigskip

\begin{center}
{\bf FIGURE CAPTIONS}

\bigskip
\end{center}

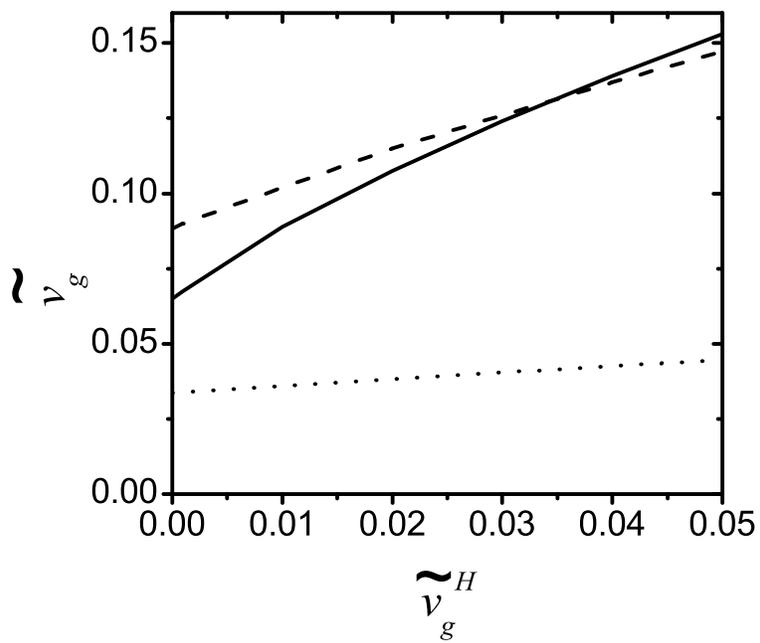
Fig. 1. Renormalized group velocity $\tilde{v}_{g}$ as a function of the
group velocity within the Hartree approximation $\tilde{v}_{g}^{H}$, in
units of $e^{2}/\pi \hbar \varepsilon $. The solid, dashed and dotted curves
correspond to $\hbar \omega _{c}/k_{B}T=10$, $15$ and $74.5$, or $T\approx
31.3$, $20.8$ and $4.2$ K respectively for $B=15.7$ T and electron density $%
n_{s}=3.8\times 10^{11}\;$cm$^{-2}$ (the $\nu =1$ quantum Hall regime is
considered)$.$

\smallskip

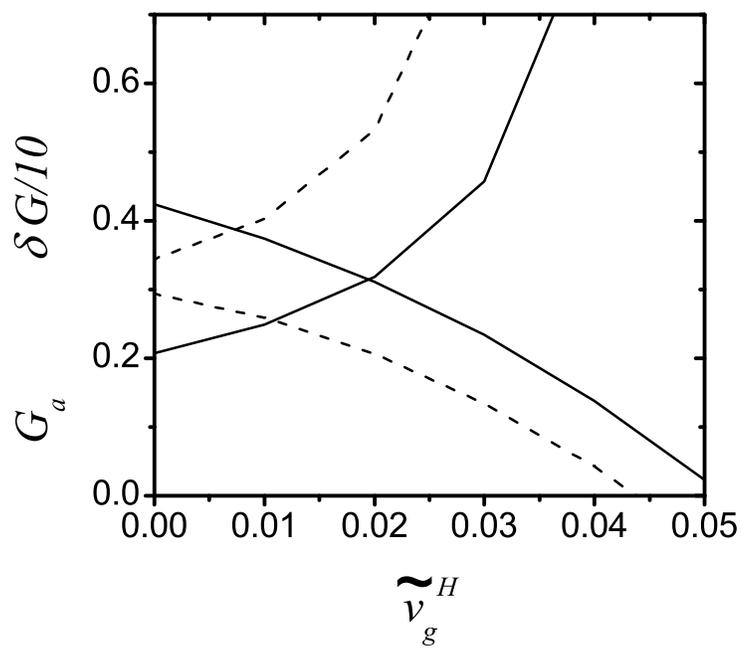
Fig. 2. Many-body enhancement of the activation gap $G_{a}=G/(\hbar \omega
_{c}/2)$ and fractional difference $\delta G=(G_{-1,1}-G)/G$, where $%
G_{-1,1}=E_{0,0,-1}-E_{0,0,1}$, as a function of $\tilde{v}_{g}^{H}$. $%
\delta G$ displays the asymmetry of the Fermi level position within the
Fermi gap in the bulk of the channel. The solid and dashed curves correspond
to the solid and dashed curves in Fig. 1. $G_{a}$ is represented by curves
with negative slope and those with positive slope corresponds to $\delta G$;
for $\omega _{c}/\Omega =50$.

\smallskip

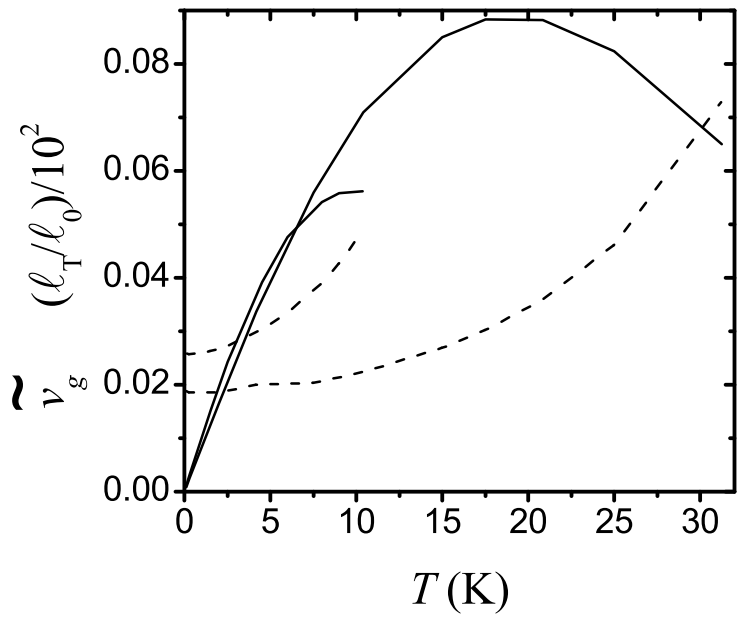
Fig. 3. Renormalized group velocity $\tilde{v}_{g}$ as a function of
temperature for $\tilde{v}_{g}^{H}=0$ is represented by solid curves; $\ell
_{T}/\ell _{0}$ is shown by dashed curves. The long curves correspond to the
sample parameters $n_{s}=3.8\times 10^{11}\;$cm$^{-2}$; $B=15.7$ T, $%
r_{0}=0.66$, $k_{\text{bulk}}(r_{0})=0.77$, while the short curves
correspond to $n_{s}=1.26\times 10^{11}\;$cm$^{-2}$; $B=5.2$ T, $r_{0}=1.15$%
, $k_{\text{bulk}}(r_{0})=0.67$. The curves are plotted only for $T\leq
\hbar \omega _{c}/10k_{B}$. For $T\alt1$ K, the upper and lower dashed
curves give $2\ell _{T}/\ell _{0}\approx 5.2$ and $3.8$, respectively.

\smallskip

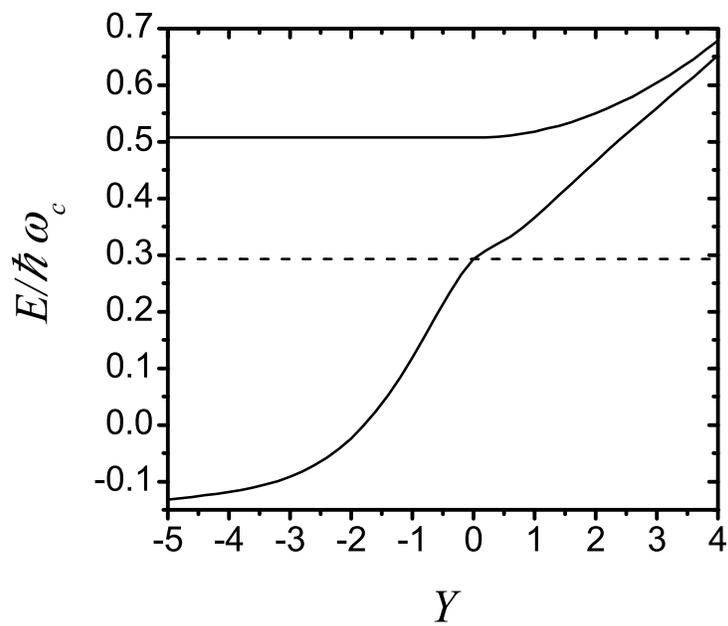
Fig. 4. Energy spectra as a function of $Y=\ell_{0}^{2}(k_{x}-k_{r0}^{(1)})/%
\ell _{T}$ for $\tilde{v}_{g}^{H}=0$ and same parameters used for the solid
curves in Figs. 1 and 2; $\hbar \omega_{c}/k_{B}T=10$, $B=15.7$ T, $\omega
_{c}/\Omega =50$ and $\tilde{v}_{g}=0.065$. Here $k_{r0}^{(1)}=k_{r}$ as $%
\tilde{v}_{g}^{H}=0$. The lower solid curve represents $E_{0,k_{x},1}$,
while the upper solid curve is $E_{0,k_{x},-1}=\varepsilon _{0,k_{x},-1}$.
The dashed horizontal line is the Fermi level of the interacting 2DES.

\end{document}